\documentstyle[twocolumn,aps,prl]{revtex}
\newcommand{\dg}{$^{\circ}$}
\begin{document}
\draft
\title{Comment on the Measurement of Cosmological Birefringence} 
\author{J. P. Leahy}
\address{University of Manchester, NRAL, 
Jodrell Bank, Macclesfield, Cheshire SK11 9DL, England}
\date{\today}
\maketitle


Nodland and Ralston [1] claim evidence for a 
wavelength-independent cosmic birefringence.
Although it has been shown [2]
that the data analysed in [1] does not
support this conclusion, the possibility of such
an effect remains interesting.
Here I present a much more accurate test.

It is well known [3]
that there is a close relation between the polarization and total
intensity structures of the double lobes of radio galaxies and quasars.
In particular the projected magnetic field (perpendicular to the
Faraday-corrected polarization orientation $\chi$)
is predominantly perpendicular to strong gradients in total intensity,
in excellent agreement with theory and MHD simulations [3,4].
These qualitative patterns are visible out to  $z > 3$ [5].
The complexity of the radio structure guarantees that the integrated $\chi$, 
an intensity-weighted 
vector average, is a small residual with only a loose relation to the
orientation of the lobes on the sky, $\psi$ 
(often mis-described as the major axis of the galaxy).
Unfortunately [1] and others use
$\chi - \psi$ as an indicator of the birefringent rotation $\beta$, 
which inevitably introduces
a great deal of noise into any relation with other quantities.

Consider instead the detailed agreement
available at high resolution, for instance Fig.~\ref{fig}.
Using the well-established trend noted above, we can determine $\beta$ 
accurately by comparing the local $\chi$ values with the local direction
of the intensity gradient, $\phi$.
A clear peak in the $\chi - \phi$ histogram is apparent at 
$-$2\dg $\pm$ 2\dg, consistent with zero cosmic
birefringence.  A major contribution to the peak comes from the
edges of the lobes, where alignment is excellent in most sources.
Table~\ref{tab} lists similar results for all 
those radio sources with $z > 0.3$ for which suitable 
digital images were to hand, together with 3C\,9, in which the
alignment between polarization and jet direction has been studied
quantitatively [6]. 
Here $\beta({\rm pred})$ gives the predicted rotation according to
the fit of [1], while $\beta({\rm obs})$ gives the observed
rotation based on the mode of the $\chi - \phi$  histogram for each
object.  
For sources from  [7] and [8], $\chi$
has been fully Faraday-corrected;
for the others, I use images at $\lambda$3.6~cm, corrected for
integrated Faraday rotation; differential effects are
at most a few degrees.
In no case is a significant $\beta$(obs) detected; any dipole along the axis
found in [1] must be $\lesssim 3$\% of the claimed amplitude.

I thank J. Ralston for constructive criticism.

\setlength{\unitlength}{1cm}
\begin{figure}
\begin{center}
\begin{picture}(8.6,5.1)
\put(0,5.7){\includegraphics{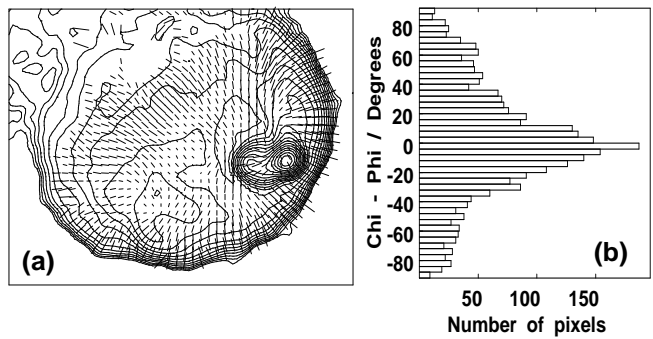}}
\end{picture}
\end{center}

\caption{(a) Contours of total intensity for the south lobe of the 
quasar 3C\,47, with line segments showing $\chi$.
(Faraday rotation in the north lobe is too large to correct).
(b) Histogram of $\chi - \phi$, 
at all points where $\sigma_{\chi - \phi} < 15$\dg.}
\label{fig}
\end{figure}

\begin{table}
\caption{Data on selected radio sources.
\label{tab}}
\begin{tabular}{ldrr@{$\pm$}rl}
Name & redshift, $z$ 
& $\beta({\rm pred})$ &\multicolumn{2}{c}{$\beta({\rm obs})$}
 & ref. \\ \tableline
3C\,9     & 2.012  &    125\dg &    2\dg &  3\dg & [6] \\
3C\,34    & 0.6897 &     54\dg &    8\dg & 12\dg & [8] \\
3C\,47\,S & 0.425  &     35\dg & $-$2\dg &  2\dg & [7] \\
3C\,55    & 0.735  &     34\dg &    6\dg &  9\dg & [9] \\
3C\,228   & 0.5524 & $-$113\dg &$-$13\dg & 20\dg & [8] \\
3C\,244.1 & 0.428  &  $-$49\dg & $-$1\dg &  9\dg & [9] \\
3C\,265   & 0.8108 &  $-$94\dg &    0\dg & 10\dg & [9] \\
3C\,268.1 & 0.97   &  $-$32\dg &   13\dg & 20\dg & [9] \\
3C\,330   & 0.428  &     14\dg & $-$5\dg & 12\dg & [9] \\
3C\,340   & 0.7754 &     49\dg & $-$6\dg &  9\dg & [8] \\
\end{tabular}
\end{table}
\clearpage
\centerline{{\large\bf Reply to astro-ph/9706126}}

\bigskip

Nodland \& Ralston's response \cite{NR97C} to my comment \cite{L97A} on
their paper \cite{NR97} points out an error (now corrected in the revised 
submission), which does not affect my conclusion. They
also accuses me, falsely, of selecting unrepresentative data, and put
forward a number of other criticisms, none of which are valid. 

As noted by \cite{NR97C}, I do not address their correlation in \cite{L97A}. 
Instead I address their conclusion, that there is cosmic birefringence, 
which was a major feature of their title, abstract, text, and media publicity.
I believe that \cite{refutes} have already
disposed of the correlation, notwithstanding the counter-arguments presented 
in \cite{NR97B} and \cite{NR97C}.

My comment, centrally, presents an operationally {\em different} measure of 
birefringence from that used by \cite{NR97}; one which is much more accurate
that the original.  Because the scatter in my measure is so small, 
data on only a few representative objects gives a much clearer result than 
statistical analysis of the large database used by \cite{NR97}. 
My measure has a clean observational and theoretical grounding (despite
the misinformed criticisms in \cite{NR97C}, discussed below), so
that any rotation detected would almost certainly be a propagation effect. 
Conversely and obviously,
any propagation effect that affects the measure used by \cite{NR97} would
also show up in my measure. The two would only be incompatible 
if, as proposed by \cite{new_mods}, the original measure was affected by an 
anisotropy in the {\em source} distribution, in which case the genuine test 
of birefringence would be my measure.

I have slightly changed the notation in the revised version of \cite{L97A}
to minimise confusion. However,
I retain the use of the symbol $\beta$ to denote the angle by which the plane 
of polarization is rotated due to cosmic birefringence, following
Eq. (1) of \cite{NR97}. On this definition $\beta$ is intrinsically
a scalar variable, although any observational estimator is subject to 
$n\pi$ ambiguity.  Eqs. (2) of \cite{NR97} resolve this by arbitrary 
{\it fiat} which frequently disagrees with the $n = 0$ required to make
sense of the data (see below), and I therefore avoid this approach.
In passing, I note that there is no sign ambiguity; 
counterclockwise rotation (looking along the ray) is positive, as standard 
in astronomy (the value of $\beta$(obs) for 3C\,47 was originally
quoted unsigned in the text of \cite{L97A}; this has now been changed).

I am grateful to Nodland \& Ralston \cite{NR97C} 
for pointing out that my calculation of 
$\beta$ from their formula was wrong (due to a programming error). 
I have revised these values and believe that they are now correct.\footnote{%
Pace \cite{NR97C}, $\cos(\gamma)$ for 3C\,47 is definitely positive,
using either of the poles determined by \cite{NR97}  
and irrespective of whether the equinox of these poles is 1950 or 2000.
For the values in Table 1, I used the first pole and assume equinox
1950, as this was used for the source positions in the references 
cited by \cite{NR97}.}. The changed values of $\beta$(pred) do not
affect the limit I can place on any birefringence.

The principle objection raised by \cite{NR97C} is that the narrow
dispersion around zero of the $\beta$(obs) values listed in Table~1
of \cite{L97A}
is due to my selection of a few special cases (or even special parts 
of special cases), and to my neglect of a large body of data that would
have given a different result. This is simply a fantasy. Given that
the numbers in the Table are representative, the other objections 
raised in \cite{NR97C} collapse.

The selection of sources, as stated in \cite{L97A}, was determined 
by availability of (digital) data to me (except for 3C\,9);
this was because a fresh analysis of the images
was required to derive the gradient angles and difference them from
the polarization angles.
These images were
obtained for a variety of purposes, unrelated to the current
topic.  I excluded objects at $z < 0.3$ since they have little predicted 
rotation according to \cite{NR97}.
I also originally excluded three unpublished
objects at moderate redshift with low $\beta$(pred). Due to the error 
in my program, two of these should have been included; given the
suspicions in \cite{NR97C}, I have included all three
in the revised version.  
I have access to no
data on any other objects at $z > 0.3$ suitable for this analysis (the
requirements for ``suitability'' are given below).
I have re-phrased the text of \cite{L97A} to explicitly state this lack of 
bias.

I did not, of course, select parts of sources because they ``look pretty''.
As clearly implied in \cite{L97A},
I used histograms of $\chi-\phi$  for each object using every point with
good signal-to-noise and for which Faraday rotation could be either 
accurately determined or reasonably neglected, as described
below.

Fig. 1a of \cite{L97A} shows a restricted region of 3C\,47
(actually most of southern lobe) because this allows a sufficiently
large scale for the polarization pattern and its relation to the total
intensity structure to be clearly visible, and because the data in the 
rest of the source does not meet the criteria noted above.
In particular, as illustrated in \cite{Leah96},
the Northern lobe is heavily depolarized at 18 cm, implying high Faraday 
depths which makes accurate determination of Faraday rotation with this data 
impossible.  3C\,47 is the only case where a substantial fraction of the 
object is obscured by the Faraday effect.

All the histograms follow the general pattern of Fig. 1b of \cite{L97A}, i.e.
they show that, while all values of $\chi-\phi$ occur, there is clearly
a preferred value, which is close to zero as expected from the alignment
effect. Thus Fig. 1 of \cite{L97A} 
does show undetectably low birefringent rotation in the direction of 
3C\,47; Table 1 establishes that negligible birefringence occurs in a 
variety of directions spread around the northern sky. 

If it were granted that the
objects I listed were not maliciously or incompetently selected,
my data alone would establish the alignment trend illustrated in Fig. 1
of \cite{L97A},
since one requires both a low birefringence and a consistent intrinsic 
alignment to obtain a small scatter of $\beta$(obs) around zero. 
However, the trend is also well documented in the authoritative
reviews I cited \cite{reviews}.

The data quality requirements for images which clearly show the
polarization structure are strict:
Firstly, the data must be good enough that accurate polarization
angles are determined in $\gtrsim$50 independent positions, 
to ensure that at least the overall structure of the the field
is resolved. Secondly, Faraday rotation should be corrected, 
ideally at each pixel using 
images at several wavelengths to extrapolate the angles to zero wavelength.
The extensive data required for this is rarely available. Alternatively,
we can apply an integrated correction (usually from published
single-dish results), providing that the differential Faraday rotation is 
small at the wavelength of observation. The latter can be checked, as large
differential rotation causes depolarization of the integrated 
emission, which is also noted in the single-dish publications. In practice
this usually means that images with $\lambda \lesssim 6$~cm are required.
This correction has been applied to all the $\lambda$3.6~cm $\beta$(obs)
values in the revised version as it makes a noticable difference (7\dg) for
the case of 3C\,55. Changes to the values reported in the original
are at most 2\dg (for 3C\,9).

Contrary to the reference in \cite{NR97C} to ``hundreds of examples
not looking like this...'', I'd be surprised if there are more than about a 
hundred published cases meeting the above requirements, nearly all being at 
low redshift, and the vast majority\footnote{%
A partial exception is the case of jets in FR\,I sources, in which the 
polarization often parallels the contours; but these are quite distinct 
in total intensity structure from the FR\,II sources analysed here.}
following the alignment trend. I note that \cite{NR97C} do not cite any
specific counter-examples, although a few surely exist.

Thus on purely observational grounds there is a high probability that the 
mode of each $\chi-\phi$ histogram is close to the value of 
$\beta$ ($\pm n \pi$).  Since in all seven (now ten) cases the mode
is close to zero,
assigning a value of $n$ other than zero (as would result in some cases from 
Eqs. 2 of \cite{NR97}) would imply a cosmic conspiracy,
in which large-angle birefringent rotations always occurs in undetectable
units of $\approx\pi$. 

The dismissal of well-established observational evidence in \cite{NR97C}
is paralleled by the dismissal of the widely accepted theoretical 
interpretation \cite{reviews} as circular reasoning.
Fortunately, we are no longer in the know-nothing phase of 35 years ago, 
when polarization was first detected. The current ``beam model'' is 
not merely consistent with the observations;  it has had several
substantial successes in  explaining, without modification, new
observations quite different in kind from
the gross total intensity images that inspired it.
One of these is its success in accounting
for polarization structure: as noted in the papers I cited \cite{theory},
practically any magnetic configuration assumed for the initial beam 
generates projected
field patterns which qualitatively resemble the observations;
in fully 3-D simulations, the agreement becomes quantitative.

\end{document}